\documentclass[aps,prl,twocolumn,showpacs,amsmath,amssymb,floatfix,superscriptaddress]{revtex4}
\usepackage{graphicx}
\newcommand {\be}{\begin{equation}}
\newcommand {\ee}{\end{equation}}

\begin{document}

\title{Nonequilibrium Invariant Measure under Heat Flow}
\date{\today}

\author{Luca Delfini}
\affiliation{Istituto dei Sistemi Complessi, Consiglio Nazionale
delle Ricerche, via Madonna del Piano 10, I-50019 Sesto Fiorentino, Italy}

\author{Stefano Lepri}
\affiliation{Istituto dei Sistemi Complessi, Consiglio Nazionale
delle Ricerche, via Madonna del Piano 10, I-50019 Sesto Fiorentino, Italy}

\author{Roberto Livi}
\altaffiliation[Also at ]{Istituto Nazionale di Fisica Nucleare,
Sezione di Firenze and INFM-CNR, Roma.}
\affiliation{Dipartimento di Fisica and CSDC, Universit\`a di Firenze,
via G. Sansone 1 I-50019, Sesto Fiorentino, Italy }

\author{Antonio Politi}
\affiliation{Istituto dei Sistemi Complessi, Consiglio Nazionale
delle Ricerche, via Madonna del Piano 10, I-50019 Sesto Fiorentino, Italy}

\begin{abstract}
We provide an explicit representation of the nonequilibrium invariant measure for
a chain of harmonic oscillators with conservative noise in the presence of
stationary heat flow. By first determining the covariance matrix, we are able
to express the measure as the product of Gaussian distributions aligned along
some collective modes that are spatially localized with power-law tails. Numerical
studies show that such a representation applies also to a purely deterministic model,
the quartic Fermi-Pasta-Ulam chain.
\end{abstract}


\maketitle

Characterizing the invariant measure of systems steadily kept out of equilibrium
is one of the main challenges towards the construction of a nonequilibrium
statistical thermodynamics. Some insight has been gained over the years mostly
thanks to the solution of stochastic models \cite{jona}. In particular, heat
conductivity in one-dimensional systems has played a prominent role, because its
understanding would also imply giving a microscopic basis to the Fourier's law
\cite{rep}. The first step dates back to 1967, when the problem of a harmonic
chain in contact with two thermal reservoirs at different temperatures was
solved exactly \cite{RLL67}. Unfortunately, the integrability of the underlying
dynamics leads to several unphysical features (e.g. a vanishing temperature
gradient) \cite{rep}. Later, purely stochastic models, where energy is assumed
to diffuse between neighbouring boxes (the oscillators), have been considered
\cite{KMP82,giardina}. More recently, systems of harmonic oscillators exchanging
energy with ``conservative" noise have been proven to admit a unique
stationary state with a constant heat flux and a linear temperature profile
\cite{BO05}. Admittedly, the leap from such class of
models to even the simplest deterministic, nonlinear ones
is still a challenge for the theory \cite{BLR00}.

In this Letter we approach the problem of describing the invariant measure in
terms of the spectral properties of the covariance matrix, i.e. at the level of
two-point correlators of the relevant dynamical variables. This procedure is
often referred to in data analysis as principal component analysis
\cite{Jolli}. It amounts to expressing the initial distribution as the product
of univariate Gaussians aligned along the eigenvectors of the covariance matrix
and whose widths are given by the corresponding eigenvalues. The decomposition
is exact for multivariate Gaussians (linear processes). In order to facilitate
the comparison with the Gibbs equilibrium measure, it is convenient to express the
covariance matrix in those variables that make it perfectly diagonal at equilibrium.
At variance with purely diffusive models, the appearance of non-diagonal terms
is crucially related to the onset of a non-zero heat flux and in particular to the
existence of anomalous transport properties. Due to the almost-diagonal
structure, the eigenvectors are localized in space.

To illustrate the approach, we first introduce a stochastic model, closely
related to the one presented in  Ref.~\cite{BBO06}, in which coupled harmonic
oscillators interact also by random pairwise collisions, each conserving both
energy and momentum.  Due to the linearity of the associated master equation,
the equations for the two-point correlators are closed and can be computed
exactly without any factorization assumption. Another useful property of the
model is that it encompasses, as a limit case, an anomalous regime, in which the
relevant transport coefficient, the thermal conductivity, diverges by virtue of
total momentum conservation \cite{NR02}. This allows comparing the cases of normal
and anomalous conductivity. In order to test the generality of these results, we
finally perform the same type of analysis by directly simulating a nonlinear
deterministic model. In spite of the unavoidable statistical fluctuations, we
find encouraging qualitative similarities.

The first model we deal with is a harmonic chain of $N$ unit-mass particles,
whose displacements and momenta are denoted by $q_i$ and $p_i= \dot q_i$.
The chain is in contact with stochastic Langevin heat baths at its extrema,
\begin{eqnarray}
\label{eq:uno}
&\dot p_i = & - kq_i + \omega^2 (q_{i+1} - 2q_i + q_{i-1}) +
\label{eq:homlin1}\\
 && \delta_{i,1}(\xi_+ - \lambda p_1) + \delta_{i,N}(\xi_- -\lambda p_N) ,
\nonumber
\end{eqnarray}
where $\xi_\pm$'s are independent Wiener processes with zero mean and variance
$2\lambda k_BT_\pm$, respectively. Fixed boundary
conditions $q_0=0,q_{N+1}=0$ are assumed. Besides satisfying dynamics
(\ref{eq:homlin1}), neighbouring particles are assumed to randomly collide, and
thereby to exchange their momenta, with a rate $\gamma$. As a result,
the phase-space probability density $P(x,t)$, that we write
in terms of the $2N$-dimensional vector whose components $x_\mu$
are $(q_1, \ldots ,q_N,p_1, \ldots, p_N)$, satisfies the master equation
\begin{equation}
\frac{\partial P}{\partial t} \;=\;
\left( \mathbb{L} + \mathbb{L}_{col} \right) P \quad.
\label{master}
\end{equation}
The first term accounts for both the deterministic force and the coupling with
the heat bath and reads
\begin{equation}
  \mathbb{L} P = \sum_{\mu,\nu}\left[
  a_{\mu\nu}\frac{\partial}{\partial x_\mu}
  (x_\nu P) + \frac{d_{\mu\nu}}{2} \frac{\partial^2 P}{\partial x_\mu
\partial x_\nu} \right]
\label{eq:liouv}
\end{equation}
where $a_{\mu\nu} $ and $d_{\mu\nu}$ are elements of the $2N\!\times\!2N$
matrices $\bf a$ and $\bf d$ that we write in terms of $N\!\times \!N$ blocks
\begin{equation}
 {\bf a} =
 \begin{pmatrix}
 {\bf 0}     &  -{\bf I}\\
 \omega^2 {\bf g}+k{\bf I}   &  \lambda {\bf r}
 \end{pmatrix}, \quad
 {\bf d} =
 \begin{pmatrix}
 {\bf 0}     &&  {\bf 0}\\
 {\bf 0}  &&  2 \lambda k_B T ({\bf r}+\eta{\bf s})
 \end{pmatrix}
\label{eq:mats}
\end{equation}
with ${\bf I}$, ${\bf 0}$ being the identity and null matrices respectively,
$r_{ij}=\delta_{ij}(\delta_{i1} + \delta_{iN})$,
$s_{ij} = \delta_{ij}(\delta_{i1} - \delta_{iN})$ and
$g_{ij} = 2\delta_{ij} -\delta_{i+1,j} - \delta_{i,j+1}$.
Moreover, we let $T=(T_++T_-)/2$ and $\eta=(T_+-T_-)/T$.
The collisional term writes
\begin{equation}
\mathbb{L}_{col} P
= \gamma \sum_i [ P(\ldots p_{i+1},p_i \ldots) -
P(\ldots p_i, p_{i+1} \ldots)]
\label{pcol}
\end{equation}
For $k=\gamma=0$ the model reduces to the one of Ref.~\cite{RLL67},
where it was showed that the nonequilibrium measure is a multivariate Gaussian
and all the second moments were exactly determined.

By denoting with $\langle . \rangle$ the average over $P$,
the covariance matrix $c_{\mu\nu}=\langle x_\mu x_\nu\rangle $
can be written as four $N\!\times\! N$ blocks,
\begin{equation}
{\bf c} =
\begin{pmatrix}
  {\bf u}       &   {\bf z}\\
  {\bf z}^\dag  &   {\bf v}
\end{pmatrix}
\label{eq:bmat}
\end{equation}
with $u_{ij} = \langle q_iq_j  \rangle$,
$v_{ij} = \langle p_i p_j \rangle$, $z_{ij} = \langle q_i p_j \rangle$;
the symbol $\dag$ denotes the transpose and we also assume
$\langle x_\mu \rangle=0$.
The evolution equation is obtained by multiplying both sides of
Eq.~(\ref{master}) by $ x_kx_l$ and thereby integrating,
\begin{equation}
\dot {\bf c} \; = \;
{\bf d} - {\bf a }{\bf c} - {\bf c}{\bf a}^\dag +\dot {\bf c}_{col} \quad.
\label{cdot}
\end{equation}
The first three terms on the r.h.s. are associated with
the operator (\ref{eq:liouv}) and are the same found in Ref.~\cite{RLL67}.
The contribution due to collisions (\ref{pcol}) reads
\begin{equation}
\dot {\bf c}_{col} = -\gamma
\begin{pmatrix}
 {\bf 0}       &   {\bf z} {\bf g} \\
{\bf g}{\bf z}^\dag  &     {\bf w}
\end{pmatrix}
\end{equation}
where the auxiliary $N\!\times \!N$ matrix ${\bf w}$ is defined by
\[
w_{ij} \equiv
\begin{cases}
v_{i+1 j} + v_{i-1 j}+ v_{i j-1} +v_{i j+1} -
4 v_{i j} & |i-j| > 1\\
v_{i\pm1 j} + v_{i j\mp1}-2 v_{ij} & i-j = \pm1 \\
v_{i-1 j-1} +v_{i+1 j+1}-2 v_{ij} & i=j
\end{cases}
\]
The matrices $\bf u$ and $\bf v$ are symmetric by construction, and one can
easily check that $\bf z$ is antisymmetric in the stationary state.

\textit{Observables -}
The most relevant observables are the kinetic temperature field
$T_i = \langle p_i^2 \rangle = v_{ii}$  and the local energy current
$J_i$, that can be written as the sum of two contributions,
$J_i = J_i^d + J_i^c$, where
\begin{eqnarray}
J_i^d &\;=\; & \omega^2 \langle q_{i-1} p_i \rangle
 \; = \; \omega^2 z_{i-1,i}
\label{deflux} \\
J_i^c &\;=\; & \frac{\gamma}{2}
[\langle p^2_{i}\rangle - \langle p^2_{i-1}\rangle ]
\; =\;  \frac{\gamma}{2} [ v_{i,i} - v_{i-1,i-1}]
\label{stoflux}
\end{eqnarray}
are the deterministic contribution (due to the springs) and the stochastic one
(due to collisions), respectively.

\textit{Coordinate change -} In the perspective of better understanding the
differences between the equilibrium and non-equilibrium invariant measure, it is
convenient to choose new variables $Q_i, P_i$ in such a way that the
equilibrium covariance matrix becomes fully diagonal. This is accomplished by the
linear trasformation
\begin{equation}
Q_i\;=\; aq_{i+1} - bq_i; \quad
P_i=p_i
\label{deltaq}
\end{equation}
with $a= [\omega^2+k/2 + (\omega^2k+k^2/4)^{1/2}]^{1/2}$, $b=\omega^2/a$.
Moreover, in analogy to Eq.~(\ref{eq:bmat}), we introduce a new covariance
matrix $\bf C$ whose components $\bf U$, $\bf V$, and $\bf Z$ are given by
$U_{ij} = \langle Q_i Q_j \rangle$,
$V_{ij} = \langle P_i P_j \rangle$, and $Z_{ij} = \langle Q_i P_j \rangle$.

\begin{figure}[h!]
\includegraphics[width=0.44\textwidth,clip]{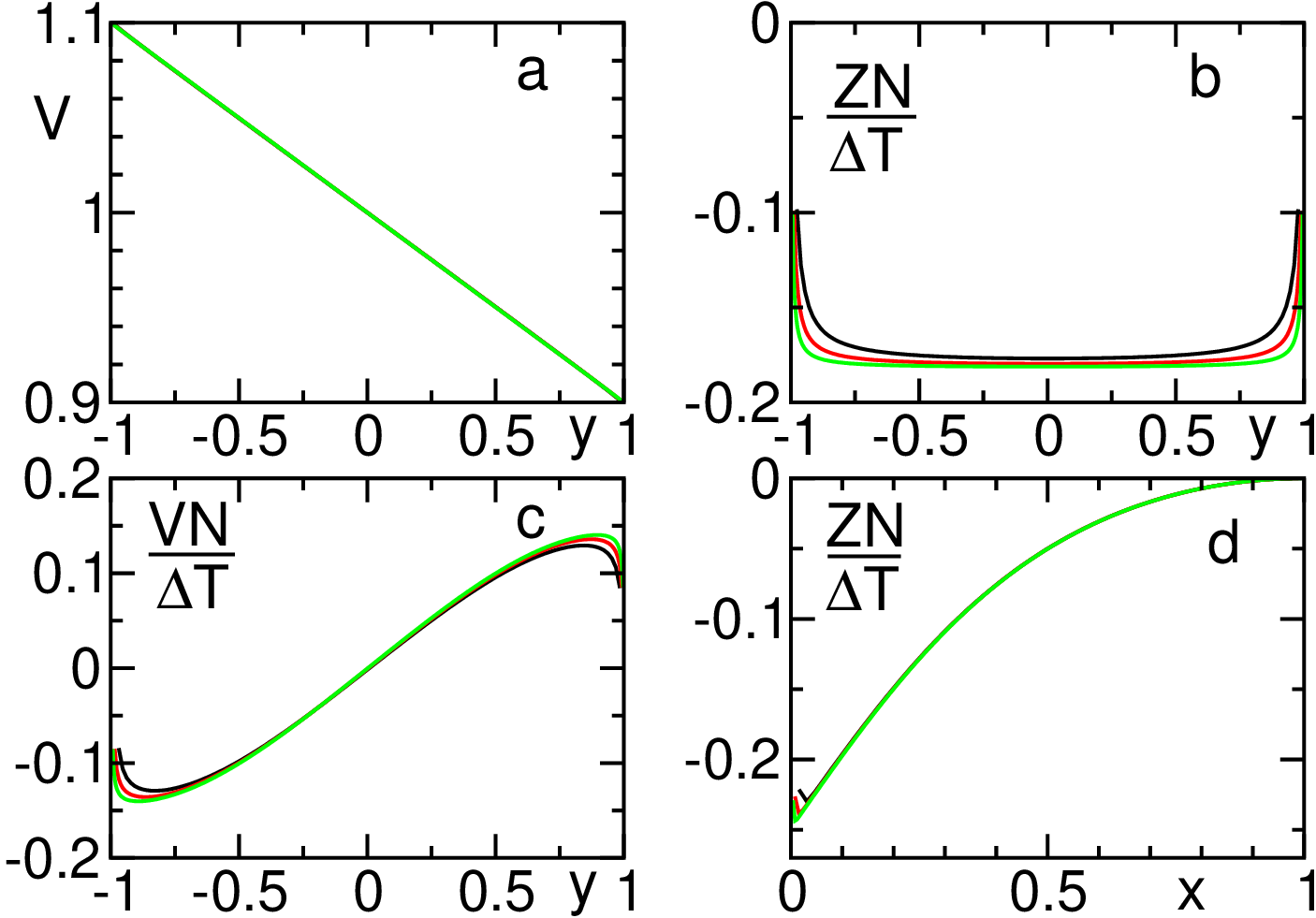}
\caption{(Color online) Some elements of the covariance matrix $\bf C$
(properly normalized)
for $k=2$, $\omega=1$, $\gamma =1 $, $\lambda=1$, $T_+=1.1$, $T_-=0.9$ and
different chain sizes $N =128,256,512$;
(a) $U_{i,i}=V_{i,i}=T_i$, (b) $NZ_{i,i+1}/\Delta T$,
(c) $NV_{i,i+1}/\Delta T$,
(d) $NZ_{i,N+1-i}$ (antidiagonal).
}
\label{fig1}
\end{figure}

\textit{Numerical solution -} The stationary solution of Eq.~(\ref{cdot})
can be efficiently computed by exploiting the sparsity of the corresponding
linear problem, as well as the symmetries of the unknowns ${\bf u}$, ${\bf v}$
and ${\bf z}$. The elements of ${\bf C}$ can be
thereby obtained from transformation (\ref{deltaq}). The diagonal elements
of $\bf C$ correspond to the temperature profile. This is illustrated in
Fig.~\ref{fig1}a, where both $V_{ii}$ and $U_{ii}$ are plotted versus
$y\equiv 2i/N -1$ for $k=2$ and different system sizes. The linear
profile confirms the validity of Fourier's law, as expected in the presence of
an on-site potential \cite{rep}. All the other $\bf C$ elements are of order
$\mathcal{O}(1/N)$ and proportional to the temperature difference $\Delta T$
and are, therefore, scaled accordingly in the other panels of Fig.~\ref{fig1}.
The upper-diagonal of $\bf Z$ is plotted in panel (b); because of the antisymmetry
of $\bf z$, $Z_{i-1,i}$ coincides with the deterministic component $J_i^d$
of the flux (see Eq.~(\ref{deflux})). The singular behaviour at the
extrema is a consequence of $J_i^d$ vanishing at the boundaries, where the
particles are not free to move. The behavior of $\bf V$ along the upper
diagonal is
presented in Fig.~\ref{fig1}c; it basically quantifies the $1/N$ finite-size
deviations exhibited by the temperature profile. Finally, the dependence of
$\bf Z$ along the principal antidiagonal [$x=(i-j)/N$] is shown in
Fig.~\ref{fig1}d, where one can notice a slow decay of the correlations. This is
analogous to what found in purely stochastic models \cite{giardina,young} and is
believed to be a generic feature of nonequilibrium stationary
states \cite{jona}.

\begin{figure}
\includegraphics[width=0.44\textwidth,clip]{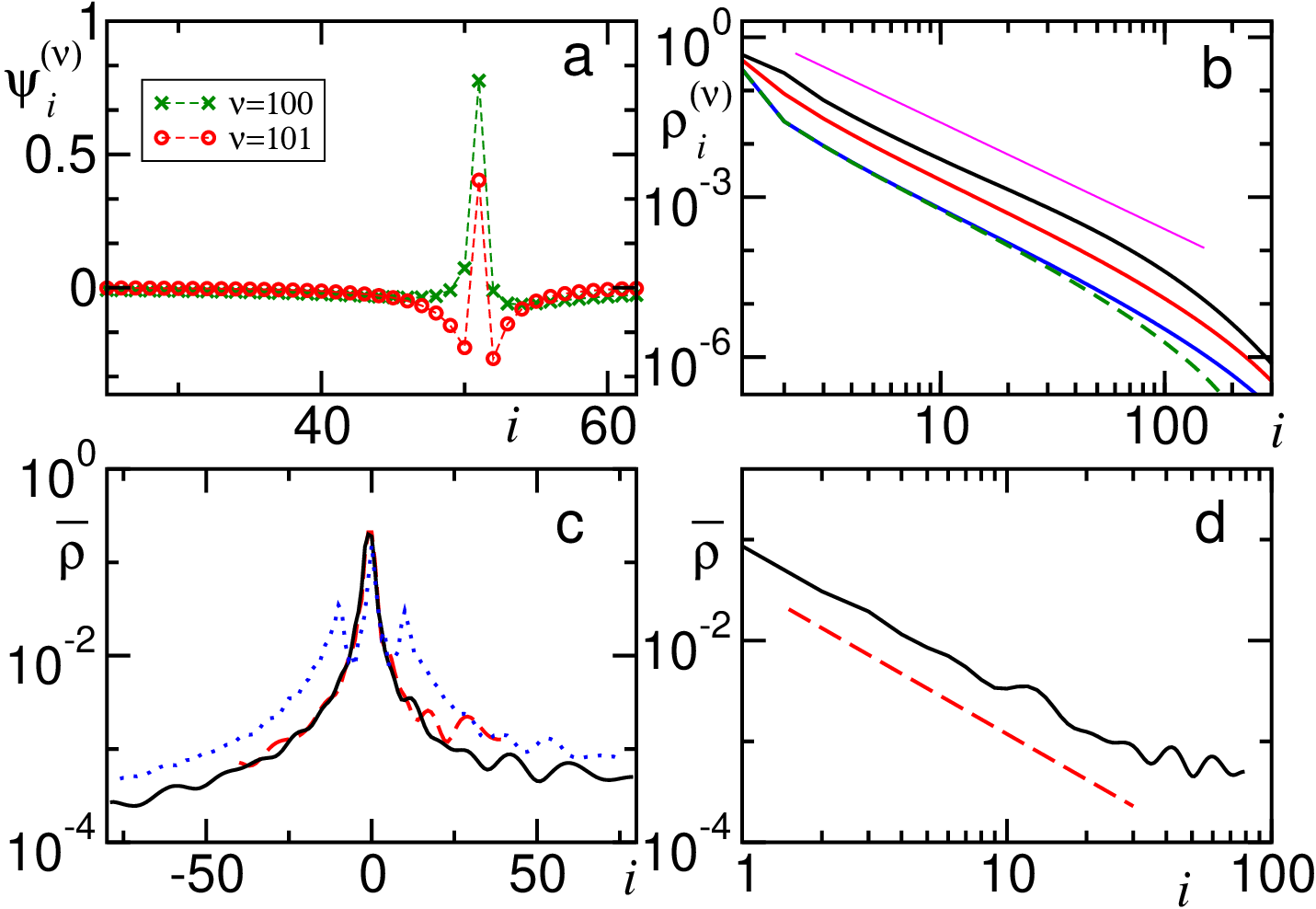}
\caption{(Color online) Localized structure of the eigenvectors of
$\bf C$. The upper panels correspond to the model and parameter values
of Fig. 1:
(a) $\psi$ components of the vectors $\nu=100, 101$
for $N=200$; (b) the squared envelope $\rho_i^{(\nu)}$ defined in the text
for $N=400$ with $\gamma=2$ (dashed line) and $N=800$ with
$\gamma = 0.5, 1, 2$ (solid lines from top to bottom).
(c) $\overline{\rho}=\sum_{\nu=\nu_1}^{\nu_2}\rho^{(\nu)}_i$: FPU model for
$N=511$, $\nu_1 =150$, $\nu_2 =350$ (dotted line), unpinned case ($\gamma=0.4$) for $N=512$, $\nu_1=150$, $\nu_2=350$ (solid line) and $N=256$,
$\nu_1=75$, $\nu_2=175$ (dashed line);
(d) same data in log-log scale for the unpinned case ($N=512$).
The straight lines have a slope $-2$ in (b) and $-1.5$
in (d).}
\label{fig2}
\end{figure}

\textit{Principal component analysis -} Once the covariance matrix is
known, it is natural to determine its eigenvalues $\lambda^{(\nu)}$
($\nu=1,\ldots,2N$) -- that we assume to be ordered from the largest to the
smallest one -- and the corresponding normalized eigenvectors, that are
denoted as
$(\phi^{(\nu)}_1,\ldots,\phi^{(\nu)}_N,
\psi^{(\nu)}_1,\ldots,\psi^{(\nu)}_N)$.
Since $\bf C$ is almost diagonal,
$\lambda^{(\nu)}$ yields the temperature profile and each of the two
eigenvector components are localized.
A less obvious fact is that the eigenvalues come in almost degenerate
pairs (the relative difference being much smaller than $1/N$) and
that the $\phi$ and $\psi$ components of corresponding
eigenvectors are localized around the same site
$m = \nu/2$ (see Fig.~\ref{fig2}a, where
the $\psi$ components are plotted for $\nu=100, 101$).
We interpret this by saying that any pair $(\nu,\nu+1)$ of eigenvectors
identifies a single, localized, degree of freedom
at equilibrium with a ``temperature" $\lambda^{(\nu)}$~.

The eigenvector spatial structure can be fruitfully illustrated by
plotting $\rho^{(\nu)}_i= [\phi_{i}^{(\nu)}]^2+[\psi_{i}^{(\nu)}]^2 +
[\phi_{i}^{(\nu+1)}]^2+[\psi_{i}^{(\nu+1)}]^2$. In Fig.~\ref{fig2}b, we see
that the vector width does not depend on the system size, which only
determines the extension of the tails. Moreover, Fig.~\ref{fig2}b shows also
that the squared envelope  decays as a power-law from its localization
center,    $\rho^{(\nu)}_i \approx [\ell/|i-m|]^{2}$ \cite{perturb}. This
defines a typical length $\ell $ that can be  interpreted as the minimal
size of the spatial region that is necessary to ensure a local thermal
equilibrium, a sort of mean free path. This interpretation  is supported
by the fact that $\ell$ decreases upon increasing  the strength $\gamma$ of
the stochastic process, which is the only thermalization mechanism
in the bulk.

\begin{figure}
\begin{center}
\includegraphics[width=0.44\textwidth,clip]{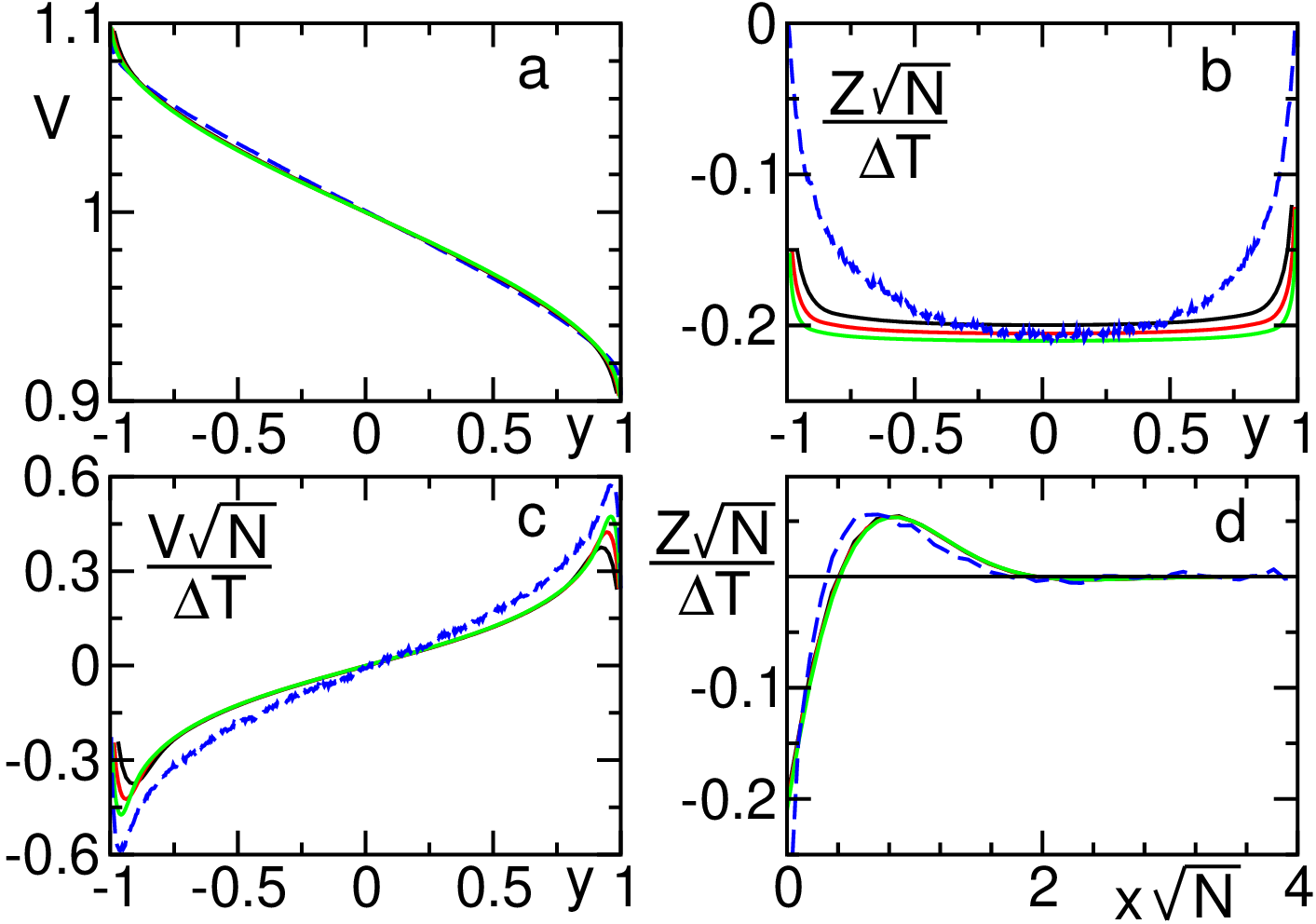}
\caption{(Color on line) The same matrix elements as in Fig.~1 in the absence
of an on-site potential
$k=0$ and for $\gamma=0.4$. The three solid lines refer to $N=128$, 256, and
512.
The dashed lines are obtained by simulation
of a chain of 511 quartic FPU oscillators.}
\label{figfpu}
\end{center}
\end{figure}

\textit{The unpinned case -} The limit case of vanishing pinning force,
$k=0$, is of particular interest. Indeed, here the total momentum is
conserved and we expect an anomalous behaviour, i.e. a diverging finite-size
thermal conductivity \cite{NR02,rep}.
As shown in
Fig.~\ref{figfpu}a
the temperature profile (solid line) is no more linear.
Remarkably, we find also that all the
nondiagonal elements are in this case of order $\mathcal{O}(1/\sqrt{N})$
(see the solid lines in Fig.~\ref{figfpu}b,c and d). In particular, through
Eq.~(\ref{deflux}) this implies that the heat conductivity diverges
as $\sqrt{N}$, in
agreement with the linear response prediction \cite{BBO06}. This observation
is supported by analytical arguments based on multiple-scale expansion of
Eq.~(\ref{cdot}) in the smallness parameter $1/\sqrt{N}$ \cite{unp}.  A
second remarkable difference with the case $k>0$ is the exponential decay of
the off-diagonal terms, the decay length scaling as $\sqrt{N}$. For
instance, in Fig.~\ref{figfpu}d (solid line) the elements along the antidiagonal
of the matrix $\bf Z$ are plotted versus $x\sqrt{N}= (i-j)/\sqrt{N}$;
similar curves are obtained for the same elements of the matrices $\bf U$ and
$\bf V$.

On the other hand, the principal component analysis reveals that the
eigenvalues still appear in almost degenerate pairs and the eigenvectors are
localized (see Fig.~\ref{fig2}c), although the tails decay with a slower power-law
exponent, close to 1.5 (see Fig.~\ref{fig2}d).

\textit{Comparison with a nonlinear model -} In order to assess the generality of
the scenario resulting from the stochastic model, it is crucial to investigate
also a nonlinear deterministic model. We have considered the Fermi-Pasta-Ulam
(FPU) chain with a purely quartic interparticle potential $(q_{i+1}-q_i)^4/4$
\cite{LLP03}. We have implemented Eq.~(\ref{eq:homlin1}), by working
with stochastic heat baths in the limit of a small mass for the reservoir's
particles
\cite{rep}. This choice allows us to use a symplectic integrator and thereby to
use ``large" time steps ($10^{-2}$), still obtaining accurate results.
The covariance matrix is then estimated by time averaging and
is thus affected by the unavoidable statistical fluctuations (we have sampled
about $5 \times 10^7$ data points separated by one time unit).
For a meaningful comparison, the $Q$ variables (\ref{deltaq}) are
chosen to be $\Omega_i(q_{i+1} - q_i)$, where $\Omega_i$ is a
renormalized frequency that can be determined by imposing that the effective
harmonic potential energy equals the kinetic energy. The homogeneity of the
potential suggests setting $\Omega_i = \zeta T_i^{1/4}$; we have found that
the best overlap between kinetic and potential energy is obtained for $\zeta =1.4$
\cite{AC03}. The results obtained
for a chain of 511 oscillators are plotted in Fig.~\ref{figfpu} (dashed lines),
where the same $\sqrt{N}$ scaling of variables as in the unpinned
stochastic model is implicitly assumed. Even though we should notice
that $\gamma=0.4$ for the stochastic model has been selected to yield the best
agreement with the temperature profile of the FPU model, the similarity between
the deterministic and the stochastic model is quite striking and extends
to the temperature deviations as well as to the behavior of the
off-diagonal terms. The only elements that are significantly different are
the upper diagonal terms of $\bf Z$ (see Fig.~\ref{figfpu}b).
Statistical fluctuations prevent an accurate analysis of the single
eigenvectors. In order to reduce their effects, we have decided to average
a subset of eigenvectors around their localization center. The results
presented in Fig.~\ref{fig2}c (the average, denoted by $\bar\rho$, 
is performed over the eigenvectors
from 150 to 350) reveal that also the eigenvectors of the FPU model
are localized (the side peaks seem to be a finite-size effect, as they tend to
disappear upon increasing the system size). A quantitative analysis of the
decay rate is out of question.

The study of the above models has shown that the key features of nonequilibrium
steady states are captured by principal component analysis and are contained in
the statement that, the invariant measure can be effectively approximated as
the product of independent Gaussians for the collective, localized mode
amplitudes. The variances are the local temperatures.

In order to further explore the validity of this claim, we have studied the
fluctuations of the single mode amplitudes and the correlations between pairs
of such modes, going beyond the second cumulants. Within the numerical accuracy,
we have not found any deviation from the assumption of a purely Gaussian
distribution, even in the FPU model. We can also conjecture that the Gaussian 
assumption should become exact in the thermodynamic limit for model 
(\ref{eq:uno})~.  We want to point out that, to
our knowledge, this is the first case where an explicit nontrivial, even if
approximate, representation of the  nonequilibrium invariant measure is given
for Hamiltonian models. As a further development, the stochastic model should
be modified to reproduce quantitatively the universal scaling laws predicted
for a generic system \cite{NR02}, as well as to analyse the corresponding
modifications on the structure of the measure.

We acknowledge useful discussions with G. Basile and S. Olla. One of us (AP)
wishes to thank the Erwin Schr\"odinger Institute for profitable exchanges of
ideas.

\end{document}